\documentclass[useAMS,usenatbib]{mn2e}
\usepackage[a4paper,totalwidth=520pt, totalheight=680pt]{geometry}

\usepackage{graphicx}
\usepackage{tabularx}

\title[Active star formation at intermediate Galactic latitude]{Active star formation at intermediate Galactic latitude: the case of IRAS~06345-3023}
\author[J. L. Yun, and P. M. Palmeirim]{J. L. Yun$^{1,2}$\thanks{E-mail: yun@oal.ul.pt} and P. M. Palmeirim$^{3}$\thanks{E-mail: pedro.palmeirim@cea.fr}\\
$^{1}$Instituto de Astrof\'{\i}sica e Ci\^encias Espaciais -- Universidade de Lisboa, Observat\'orio Astron\'omico de Lisboa, Tapada da Ajuda, \\ 1349-018 Lisboa, Portugal\\
$^{2}$Departament d’Astronomia i Meteorologia, Institut de Ci\`encias del Cosmos, Universitat de Barcelona, IEEC-UB, Mart\'{\i} i Franqu\`es 1, \\ E-08028 Barcelona, Spain\\
$^{3}$Laboratoire AIM, CEA/DSM-CNRS-Universit\'e Paris Diderot, IRFU/Service d'Astrophysique, C.E. Saclay, Orme des Merisiers, \\ F-91191 Gif-sur-Yvette, France}

\begin{document}

\date{ Accepted 2015 January 20. Received 2014 December 15; in original form 2014 December 15
}

\pagerange{\pageref{firstpage}--\pageref{lastpage}} \pubyear{2015}

\maketitle

\label{firstpage}

\begin{abstract}
We report the discovery of a small aggregate of young stars seen in high-resolution, deep near-infrared ($JHK_S$) images towards IRAS~06345-3023 in the outer Galaxy and well below the mid-plane of the Galactic disc. The group of young stars is likely to be composed of low-mass stars, mostly Class~I young stellar objects. The stars are seen towards a molecular cloud whose CO map peaks at the location of the IRAS source. The near-infrared images reveal, additionally, the presence of nebular emission with rich morphological features, including arcs in the vicinity of embedded stars, wisps and bright rims of a butterfly-shaped dark cloud. The location of this molecular cloud as a new star formation site well below the Galactic plane in the outer Galaxy indicates that active star formation is taking place at vertical distances larger than those typical of the (thin) disc. 

\end{abstract}


\section{Introduction}

Star formation occurs across the Galactic disc with most molecular material concentrated at low  latitudes $|b| < 2^{\circ}$. Both in the inner and in the outer Galaxy, young stars still partly embedded in the dense gas and dust in molecular clouds have been found \citep[e.g.,][]{tapia91, strom93, mccaughrean94, horner97, luhman98,yun09}. They represent current active star formation sites. 

The study and census of star formation sites in regions of higher Galactic latitudes have received comparatively little attention and coverage. However, the detection and characterisation of star formation sites across the whole Galaxy have strong implications on its structure and evolution. 

In our previous studies of star formation in the outer Galaxy, we have reported discovery of what can be designated as star formation sites in extreme environments, namely near the distant ``edge'' of the Galactic disc (at Galactocentric distances of about 16.5~kpc \citep{santos00}), or far below the midplane of the Galactic disc (at vertical distances of about 500~pc  \citep{palmeirim10}). In both cases, the presence of molecular material and of clusters of embedded stars in these environments posit that the physical conditions required to form a cluster of stars are met in locations at distances away from the Galactic plane (or away from the Galactic centre) larger than reasonably expected. 

IRAS~06345-3023 is an IRAS PSC source in the outer Galaxy that currently appears classified in the SIMBAD data base as `` 6dFGS gJ063630.0-302542 -- Galaxy''. This means that it is listed in the 6dFGS, the Six-Degree Field Galaxy Survey catalogue as being an extragalactic source. However, a search in the 6dFGS catalogue, at the IRAS coordinates, does not return this source. Instead, a different source (g0636403-302842) is presented, which is located about 4 arcminutes away. In addition, NED (the NASA Extragalactic Database) does not contain this source.

\citet{Gyulbudaghian} first called attention on the presence of two nebulous ``stars'' on ESO/SRC J plates at this location. \citet{Yonekura} used the 4~metre NANTEN milimetre telescope to study a sample of intermediate-to-high latitude IRAS sources and reported detection of CO emission from IRAS~06345-3023. Their CO maps, although at a coarse angular resolution, show a clear peak at the position of the source. 

As part of our study of young embedded clusters in the outer Galaxy \citep[e.g.][]{yun09,palmeirim10},
we have conducted observations (near-infrared $JHK_S$ imaging, and millimetre CO line) towards IRAS~06345-3023. These observations revealed the presence of young stars embedded in a molecular cloud core. 

We report here our near-infrared discovery of an aggregate of young stars seen towards IRAS~06345-3023. In addition, we use new CO data, to characterise the molecular environment and derive its kinematic distance. 
Section~2 describes the observations and data reduction. In Section~3, we present and discuss the results. A summary is given in Section~4.


\section[]{Observations and data reduction}

\subsection{Near-infrared observations}

Near-infrared ($J$, $H$ and $K_S$) images were obtained on 2000 November 12 and on 2002 January 8 using the ESO Antu (VLT Unit 1) telescope equipped with the short-wavelength arm (Hawaii Rockwell) of the ISAAC instrument. The ISAAC camera \citep{moorwood98} contains a 1024 $\times$ 1024 pixel near-infrared array and was used with a plate scale of 0.147 arcsec/pixel resulting in a field of view of 2.5~$\times$~2.5 arcmin$^2$ on the sky.
For each filter, dithered sky positions were observed. A series of 12 images with individual on-source integration time of 6 and 3.55 seconds was taken in the $J$ and in the $H$ bands, respectively. Similarly, series of 6 images, each of 2 second integration time, were obtained in the $K_S$ band. 

The images were reduced with the Image Reduction and Analysis Facility (IRAF), using a set of our own scripts to correct for bad pixels, subtract the sky background, and flatfield the images. Dome flats were used to correct for the pixel-to-pixel variations of the response. The selected images were then aligned, shifted, trimmed, and co-added to produce a final mosaic image for each band $JHK_S$. 
Correction for bad pixels was made while constructing the final mosaics that cover about $2.5 \times 2.5$ arcmin$^2$ on the sky.

Point sources were extracted using {\tt daofind} with a detection threshold of 
5$\sigma$. The images were inspected to look for false detections that had been included by {\tt daofind} in the list of detected sources. These sources were eliminated from the source list. Aperture 
photometry was made with a small aperture (radius = 0.5$''$) and aperture corrections, 
found from bright and isolated stars in each image, were used to correct 
for the flux lost in the wings of the PSF. The error in determining 
the aperture correction was $<$ 0.03 mag in all cases. 

We used the 2MASS All-Sky Release Point Source Catalogue \citep{skrutskie06,cutri03} to calibrate our observations.
The $JHK_S$ zeropoints were determined using 2MASS stars brighter than $K_S$ = 14.7 mag.
The standard deviations of the offsets between ISAAC and 2MASS photometry
are 0.1 mag, in all bands. Given the relatively small number of stars in these images, it is not possible to make a robust estimate of the completeness limit of the observations based on the statistics of the stars detected. However, a relatively good determination of the completeness limit can be obtained calculating {\it e.g.} the 
$5\sigma$-limit in each band. Those are 20.6 in $J$, 20.0 in $H$, and 18.9 mag in $K_S$.

\subsection{Millimetre line observations}
As part of a survey of molecular gas in the third Galactic quadrant, the region around the position of the IRAS source was mapped using the single-dish 15-m SEST radiotelescope in 1999 September. The map was obtained in the rotational line of $^{12}$CO(1-0) (115.271~GHz) and consisted of $3 \times 3$ positions, in full-beam spacing ($46\arcsec$), and centred on the IRAS coordinates. The integration time was 60 seconds. 
The spectra were taken in frequency switching mode, recommended to save observational time when mapping extended sources.
One additional central spectrum was taken in beam-switching mode, with integration time of 120 seconds, in order to obtain improved baselines to search for the presence of molecular outflows.

A high-resolution 2000 channel acousto-optical spectrometer was used as a
back end, with a total bandwidth of 86~MHz and a channel width of 43~kHz, which at the frequency of 115 GHz corresponds to approximately 0.11~km~s$^{-1}$.
The antenna temperature was calibrated with the standard chopper wheel method.
Pointing was checked regularly towards known circumstellar SiO masers, 
and pointing accuracy was estimated to be better than $5\arcsec$.

The data reduction pipeline was composed of the following steps: 
\textit{i)} folding the frequency-switched spectrum; 
\textit{ii)} fitting the baseline by a polynomial and subtracting it; 
\textit{iii)} obtaining the main-beam temperature $T_{\mathrm{MB}}$ by dividing the antenna temperature $T_{\mathrm{A}}$ by the $\eta_{\mathrm{MB}}$ factor, equal to 0.7.

The spectrum baseline RMS noise (in $T_{\mathrm{MB}}$), averaged over 
all map positions, has been found to be 0.15~K.

\begin{figure}
\centering
   \includegraphics[width=7.7cm]{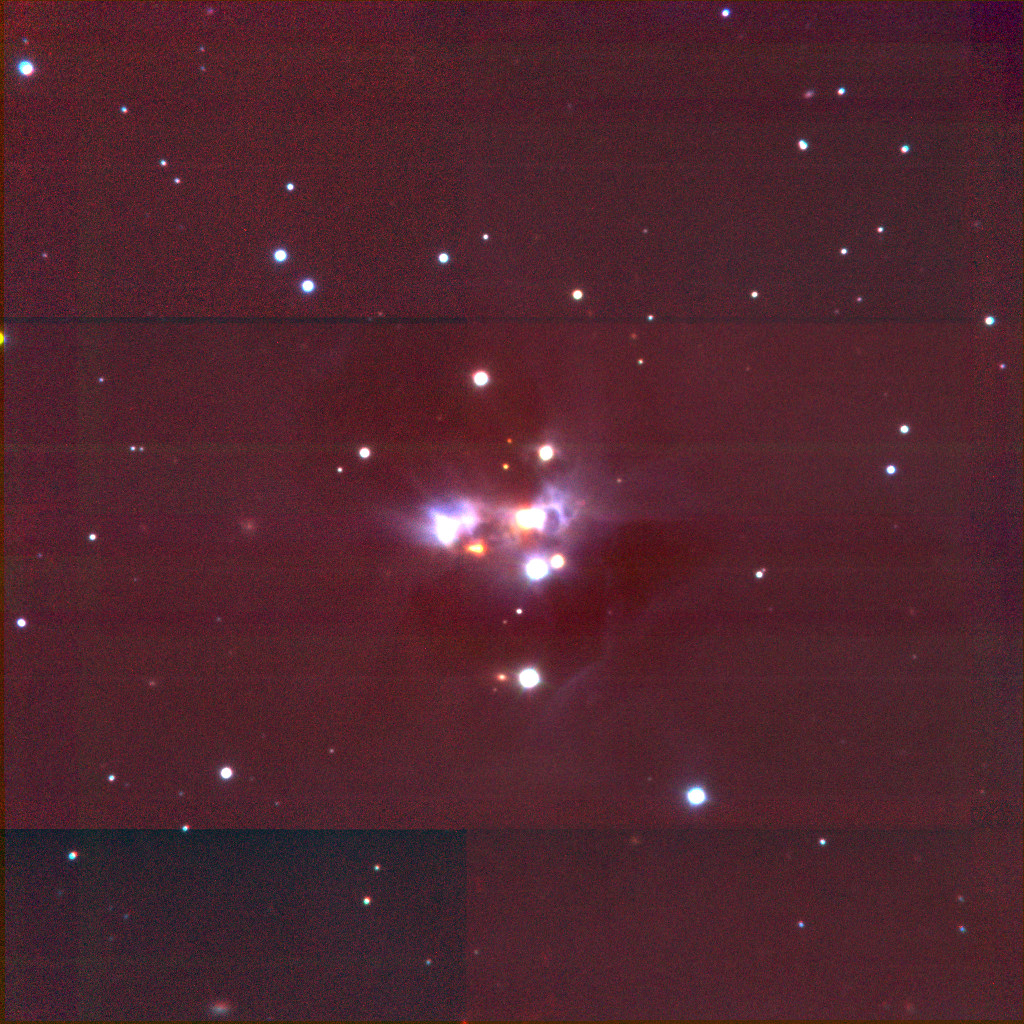}
\caption{$J$ (blue), $H$ (green), and $K_S$ (red) colour composite image towards IRAS~06345-3023 covering $2'.5 \times 2'.5$. North is up and East to the left. Notice the nebular emission around a small concentration of red stars towards the centre of the image, and the very small number of field stars.}
   \label{RGB995}
\end{figure}

\section{Results and discussion}

\begin{figure*}
\centering
   \includegraphics[width=12cm]{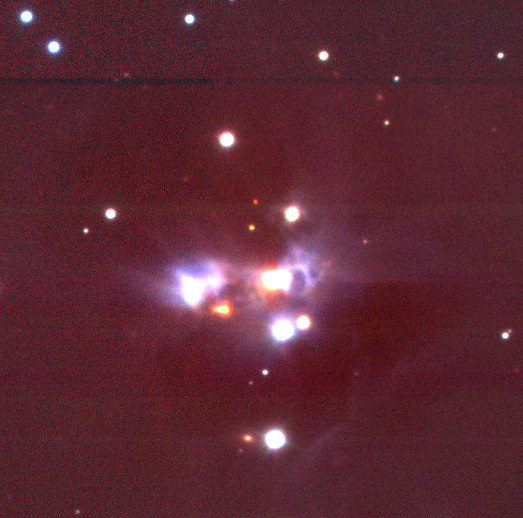}
\caption{Zoomed-in $J$ (blue), $H$ (green), and $K_S$ (red) colour composite image covering $1'.25 \times 1'.25$. North is up and East to the left. Notice the rich structure of the emission seen, including arcs, wisps and bright rims.}
   \label{RGB995-1}
\end{figure*}

\subsection{The near-infrared view}

Figure~\ref{RGB995} presents the VLT/ISAAC $JHK_S$ near-infrared colour composite image obtained towards IRAS~06345-3023. Centred on the image, extended nebular emission involving a few bright stars is displayed. In addition, when compared to other fields, usually at low Galactic latitudes, we notice a very small number of field stars, either background or foreground, that populate the Galactic disc and frequently contaminate deep near-infrared images.

\begin{figure}
   \centering
   \includegraphics[width=8cm]{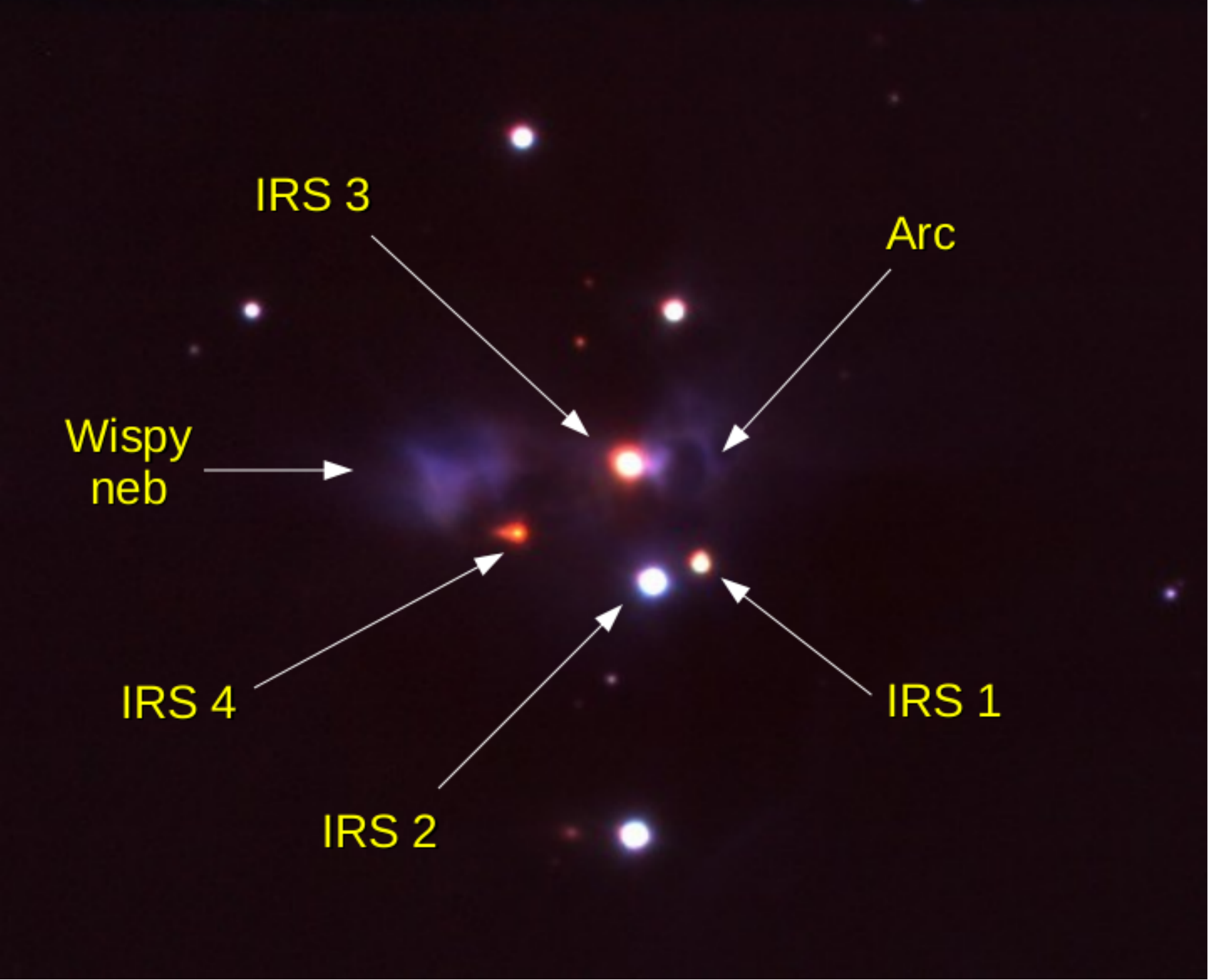}
   \caption{The same as Fig.~\ref{RGB995-1} but with a colour coding emphasizing the emission from bright stars. Notice the nomenclature of the structures labeled here, some of which can be seen more clearly in Fig.\ref{RGB995-1}} 
	\label{RGB999}%
    \end{figure}

A close look at the image (see the zoomed-in view of the same region shown in Figure~\ref{RGB995-1}) reveals the presence of a butterfly-shaped bipolar dark cloud, to the north-northeast and to the south-southwest. Notice also the bright rims of the dark ``butterfly wings'' illuminated by the central stars. The richness of the structures seen includes a quasi-circular bright ring and bright wisps or rays of light against the black opaque northern wing (see also Fig.~\ref{RGB999} for the nomenclature used here).

In Fig.~\ref{RGB999} we reproduced the colour composite image using a colour coding emphasizing the bright stars as opposed to the diffuse nebular emission. 
A few important points can be derived as follows.

The bright source labeled ``wispy neb'' is not a point source, it corresponds to pure diffuse emission. Several wisps in this nebula (better distiguished in  Fig.~\ref{RGB995-1}) seem to point away from a common point, the location of one of the reddest sources, that labeled ``IRS~4''. The bluer colour of the wispy nebula supports the idea that it corresponds to scattered light from a cavity located close to the centre of the butterfly-shaped core where opacity is lower. In fact, the position of one of the two ``nebulous stars'' that were noticed on the optical ESO/SRC images \citep{Gyulbudaghian}  coincides with the position of this wispy nebula.

The other optical ``nebulous star'' corresponds to the blue emission seen to the west including the circular arc (the ``Arc''). 
This arc-like structure appears to be associated with the source labeled ``IRS~3'' which is the brightest red embedded source present in the images. The arc is quasi-circular, very well delineated, and constitutes an engimatic structure, similar to the structure seen around a few other young stellar objects. Examples of these are the diamond-ring object of \citet{yun07}, which was conjectured to represent a circumbinary structure existing during a brief stage in the formation of a binary star; and GG~Tau, a multiple young source, which exhibits a circumbinary ring \citep[e.g.][]{pietu11,beck12}, and that has recently been suggested to be forming planets \citep{dutrey14}. IRS~3 could be another case of a multiple young object denouncing its multiplicity through the presence of circumbinary arcs or rings, seen in the near-infrared or submillimetre.

Interestingly, neither of the two optical ``nebulous stars'' is a real star (point source) but instead corresponds to diffuse scattered emission escaping through lines-of-sight of lower optical depth.
Several cavities excavated by winds from young stars are now known and have been modelled \citep[e.g.][]{whitney03,stark06} . Our images seem to confirm the rule that young stars disrupt the parent cloud core reducing the opacity from specific lines-of-sight and resulting in very non-homogenous distributions of extinction across cloud cores.
Also remarkable is the detection of wisps of bluer light escaping from this star formation site, similar to sunlight under low-lying clouds on a partly cloudy day on Earth. In addition, despite its faintness, a careful look reveals the presence of a bright rim of this molecular cloud core specially in the southwest part of the dark nebula (Fig.~\ref{RGB995-1}).

%

\subsection{Molecular gas and location in the Galaxy}

We have detected CO emission towards the positions of the nebular emission and of this group of stars. Moreover, the location of the nebula coincides with the position where the molecular gas peaks. 

\begin{figure}
   \centering
  \includegraphics[width=8cm]{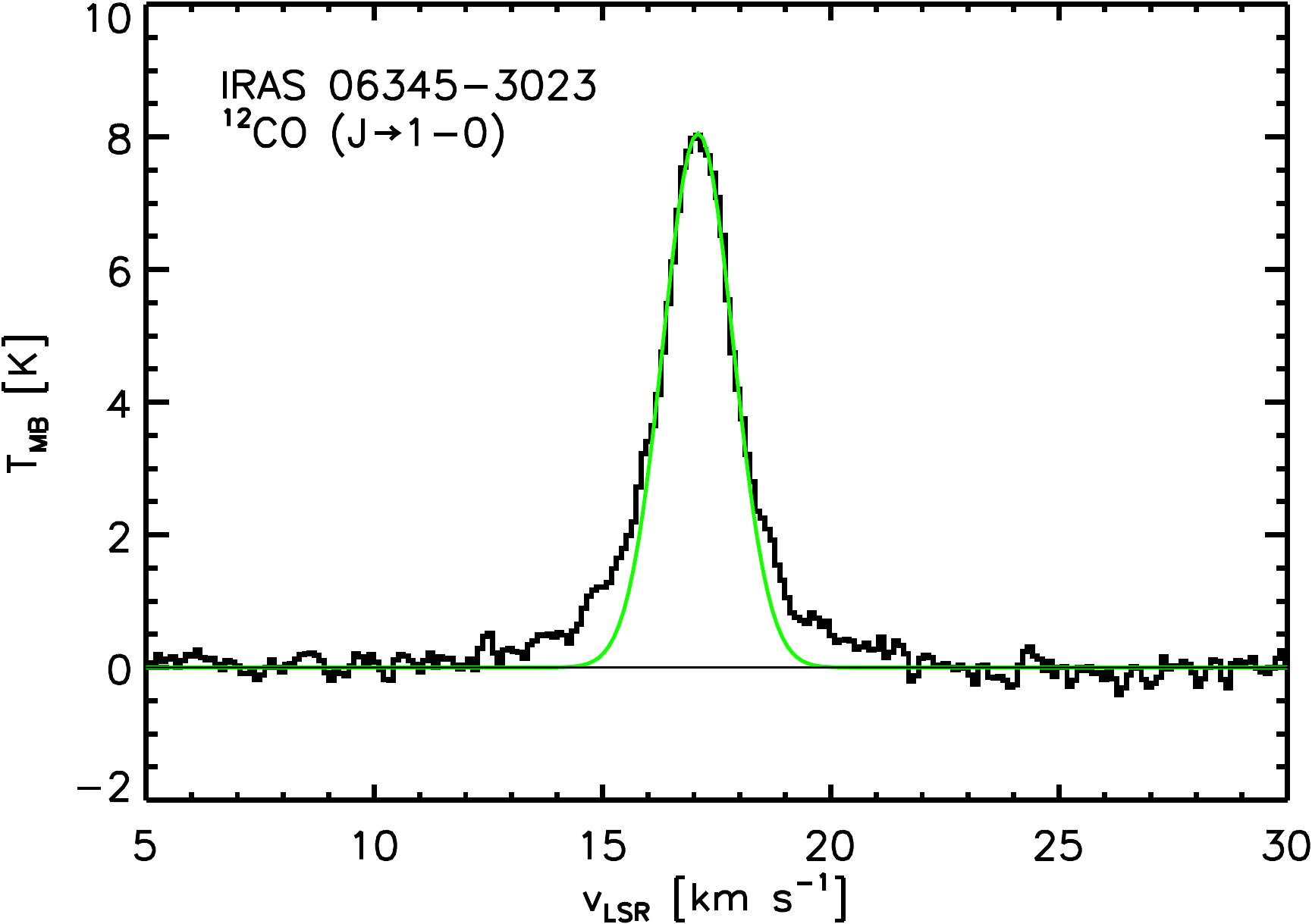}
   \caption{CO spectrum towards IRAS~06345-3023. The line deviates from a gaussian (green curve).}
             \label{co1}%
     \end{figure}

Fig.~\ref{co1} shows the beam switching spectrum from the IRAS~06345--3023 revealing a clear detection of a CO line. A gaussian fit yields a $T_{MB}$ peak value of 8.1~K at the $V_{LSR}=17.1$ km~s$^{-1}$ in good agreement with \citet{Yonekura}.
The line presents some deviation from a gaussian, exhibiting moderate wings that could be due to a molecular outflow.

\begin{figure}
   \centering
  \includegraphics[width=8cm]{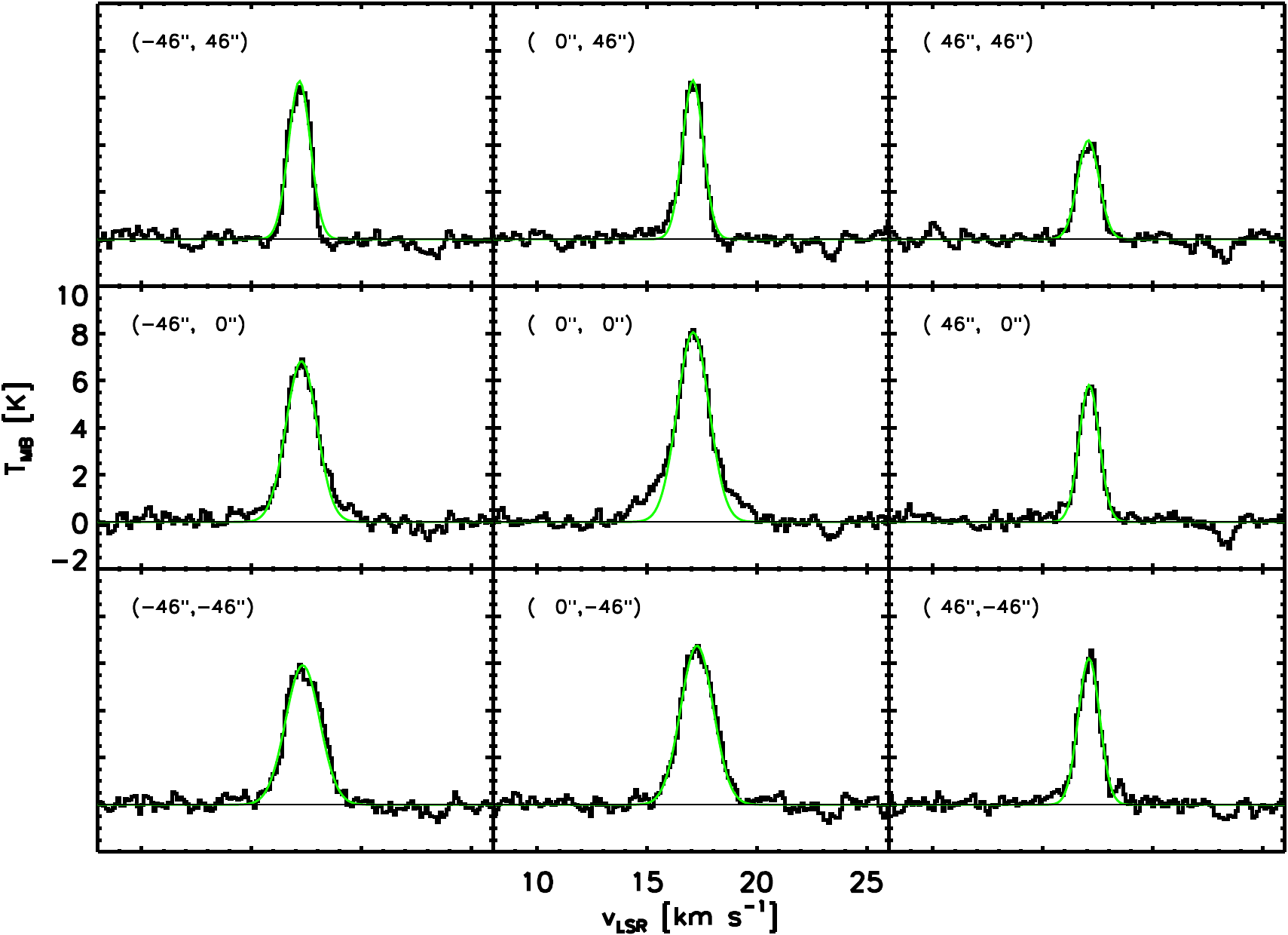}
   \caption{CO map centred at IRAS~06345-3023. The line is stronger at the centre position corresponding to the location of the infrared nebula and embedded stars.}
             \label{comap}%
     \end{figure}

Our $3\times 3$ CO map (Fig.~\ref{comap}) shows relatively constant CO emission, but peaking at the centre position where the IRAS source is located. Because the map was obtained in frequency-switching mode, we did not try to look for the presence of outflows in the beams of this map.

Either adopting a flat rotation curve in the outer Galaxy or using a circular rotation model \citep[e.g.,][]  {brand93}, a heliocentric distance $d_H=1.5$~kpc, and a galactocentric distance $d_G=10.0$~kpc are found, respectively. This is in good agreement with the distance quoted by \citet{Yonekura}.
Thus, the projected size of the region occupied by the near-infrared nebula and butterfly-wings turns out to be $\sim 0.45$~pc, a typical size for a molecular cloud core, and coincident with the C$^{18}$O-derived size \citep{Yonekura}. However, as expected, the size of the molecular cloud is much larger as can be seen by the large scale map of \citet{Yonekura} which covers more than $10'$ on the sky.

Interestingly, due to its relatively high Galactic latitude ($b=-16^{\circ}\!\!.3$), this source is located about 450~pc below the mid-plane of the Galactic disc. This value is among the largest known for molecular clouds at this distance in the outer Galaxy \citep{may97}, placing this source at the edge of the Galactic (thin) disc \citep{kalberla14}.

A similar large vertical distance was found for the star formation site reported by \citet{palmeirim10}. In that case, flaring and the location of the Galactic warp could account for the presence of the star formation site. However, this is not the case for IRAS~06345-3023. In fact, flaring of the gas component (both HI and H$_2$) in the Galactic disc is known to occur but is significant only at larger Galactocentric distances \citep{kalberla09,may97,wouterloot90}. 
Thus, the present day existence of molecular gas and dust at this relatively large vertical distance raises new questions about the evolution of the Galactic disc, as well as the ability to form stars in this extreme environment.

\subsection{Young stars}

In order to determine the nature of the group of stars seen in the near-infrared images, and specially those seen towards the nebula, photometry of the stars in the $JHK_S$ images was performed. Not many stars are seen, neither field stars (background or foreground) nor embedded young stars. The scarcity of field stars is due to the location of this star formation site in the outer Galaxy, and about 450~pc below the Galactic disc. On the other hand, given the good resolution and depth of the images, and considering the distance, the scarcity of young stars forming in this cloud core could indicate a lack of ability to form rich clusters at this large distance below the Galactic disc. 

Table~1 presents the results of our photometry for the sources detected in the $JHK_S$-bands in the region shown in Fig.\ref{RGB995-1}.
Columns (1) and (2) give their coordinates. The $JHK_S$ magnitudes (in the 2MASS system) are presented in columns (3), (4) and (5), where errors are of the order of 0.10 mag. In column (6) we indicate the sources that are likely to be young stellar objects according to their near-infrared $(J-H)$ and $(H-K_S)$ colours \citep{adams87}. The sources labeled in Fig.~\ref{RGB999} are indicated by their names. 

Only nine stars are detected as young stellar object candidates and likely to be forming in this region. Their near-infrared colours are compatible with Class~I YSOs, that is optically invisible objects still surrounded by cirumstellar discs and envelopes and frequently exhibiting modest outflows. In fact, these young stars are not optically visible (absent in the Digitized Sky Survey) even though they have created regions of lower extinction through the cloud that manifest as optical nebulae. 
We estimate the luminosity of the brightest one (IRS~3) by integrating the near-infrared fluxes and the IRAS fluxes. The result is 41$L_{\odot}$ and it represents an upper limit for the luminosity of the brightest object in the $K_S$-band. This low value of the luminosity for a pre-main-sequence object implies that no high-mass star is present in this star formation site.

\begin{table}
\caption{VLT/ISAAC Photometry of sources towards IRAS~06345--3023} 
\label{table:1}      
\centering                          
\begin{tabular}{c c c c c c c }    
\hline\hline                 
ID & R.A. & Dec & m${_J}$ & m${_H}$ & m$_{K\!s}$ & YSO  \\
& {\small (2000)} & {\small (2000)} &  &  & &  \\
\hline                        
\noalign{\vspace{2 pt}}
1 & 06 36  26.5   &-$\,$30 25 50   & 19.95 & 18.90 & 18.47 &     \\
2 & 06 36  26.6   &-$\,$30 25 51   & 17.86 & 16.94 & 16.65 &    \\
3 & 06 36  26.6   &-$\,$30 25 10   & 18.65 & 17.42 & 16.77 &    \\
4 & 06 36  27.8   &-$\,$30 25 13   & 19.25 & 18.17 & 17.96 &     \\
5 & 06 36  27.9   &-$\,$30 25 20   & 19.83 & 18.39 & 17.86 &     \\
6 & 06 36  28.6   &-$\,$30 25 10   & 16.99 & 15.56 & 14.58 &   Yes \\
7 & 06 36  28.9   &-$\,$30 25 49   & 16.49 & 14.43 & 13.28 &  Yes (IRS 1) \\
8 & 06 36  29.0   &-$\,$30 25 33   & 15.53 & 13.95 & 12.80 &   Yes \\
9 & 06 36  29.1   &-$\,$30 25 50   & 13.43 & 12.46 & 11.81 &  Yes (IRS 2)  \\
10 & 06 36  29.2   &-$\,$30 26 06   & 13.53 & 12.48 & 11.98 &     \\
11 & 06 36  29.2   &-$\,$30 25 43   & 15.31 & 13.17 & 11.19 &  Yes (IRS 3) \\
12 & 06 36  29.3   &-$\,$30 25 56   & 18.80 & 17.58 & 17.16 &    \\
13 & 06 36  29.4   &-$\,$30 25 35   & 20.32 & 18.07 & 16.89 &   Yes \\
14 & 06 36  29.5   &-$\,$30 26 06   & 19.41 & 17.94 & 17.04 &   Yes \\
15 & 06 36  29.7   &-$\,$30 25 22   & 14.83 & 13.70 & 12.88 &   Yes \\
16 & 06 36  29.8   &-$\,$30 25 47   & 19.36 & 16.28 & 13.55 &  Yes (IRS 4) \\
17 & 06 36  30.2   &-$\,$30 25 04   & 16.33 & 15.47 & 15.12 &    \\
18 & 06 36  31.0   &-$\,$30 25 33   & 16.21 & 15.17 & 14.84 &  \\
19 & 06 36  31.3   &-$\,$30 25 35   & 19.29 & 18.10 & 17.68 &   \\
20 & 06 36  31.7   &-$\,$30 25 08   & 15.07 & 14.57 & 14.39 &   \\
21 & 06 36  32.0   &-$\,$30 25 04   & 15.22 & 14.61 & 14.38 &   \\
\hline\hline                                
\end{tabular}
\end{table}

\subsubsection{WISE counterparts}

In order to search for further information about these VLT/ISAAC sources, we acessed the archived database of the Wide-field Infrared Survey Explorer (WISE) satellite \citep{wright10}. 
WISE observed in the mid-infrared 3.4, 4.6, 12.0, and 22.2 $\mu$m. The beam sizes are 6.1$''$, 6.4$''$, 6.5$''$, and 12.0$''$, respectively.

Spectral indices of candidate YSOs, contained in the WISE All-Sky Source Catalog for the point sources, calculated between 3.4 and 22 $\mu$m (or between the $K_S$-band and the $[22]$ WISE band), can be used to yield a YSO classification \citep{wilking01, liu11}, allowing to distinguish Class~I YSOs from ``flat spectrum'' YSOs, and from Class~II YSOs. 

Inspection of a multi-colour image (composite of WISE bands images) centred at IRAS~06345--3023  reveals that this star formation site is located at the northeastern edge of an extended molecular cloud, traced by the 12 and 22 $\mu$m WISE emission, and the IRAS IRIS 100 image, in good agreement with the coarse CO map of \citet{Yonekura}.
In addition, within $10'\times 10'$, there are only a few red sources which are located at the centre of the image coincident with the position of the $JHK_S$-derived YSOs. No other red sources were detected by WISE within this field, reinforcing the idea that no additional young stars exist in this star formation site. 

The angular separations between stars in the region containing IRS~1, IRS~2, IRS~3, and IRS~4 are of the order of $6''$, a value similar to the WISE angular resolution. Thus, not surpringly only one WISE source is detected in this region. The location of this WISE source is closest to IRS~3, the brightest of these four YSO-candidates. This blending prevents us from using the WISE magnitudes to classify these YSOs individually. We opted to assign this WISE source to IRS~3.  Furthermore, of the remaining ISAAC sources listed in Table~1, only four have WISE counterparts. Those are listed in Table~2. Calculation of their WISE spectral indices (between 3.4 and 22 $\mu$m) led to the YSO classification given in Table~2 (the classification remains un-changed if we use the spectral indices between $K_S$ and $[22]$ instead).

\begin{table*}
\begin{minipage}[t]{\columnwidth}
 \caption{ISAAC sources towards IRAS~06345--3023 with WISE counterparts} 
\label{table:2}      
\centering                          
\renewcommand{\footnoterule}{}  
\begin{tabular}{c c c c c c c}    
\hline\hline                 
ID & WISE & ${[3.4]}$ & ${[4.6]}$ & ${[12]}$ & ${[22]}$ & WISE  \\
  & name &  &  & & & YSO class \\
\hline                        
\noalign{\vspace{2 pt}}
5  &  J063627.98-302520.0  & 15.313   & 14.558 & 9.521 & 6.427 &  Class I   \\
6   & J063628.61-302509.1  & 13.523 & 12.822 & 10.073 &  7.105 & Flat spectrum \\
10   & J063629.17-302605.5   & 10.780 & 10.108 & 7.991 & 4.600   & Flat spectrum \\
11   & J063629.28-302543.4   & 9.148 & 7.887 & 5.310 &  2.375  &  Class I\\
15   & J063629.71-302521.5  & 11.994 & 11.373 & 9.154 & 4.933 & Class I \\
18   & J063630.95-302533.1   & 14.149 & 13.448 & 11.432 & 5.192 & Class I \\
\hline\hline
\end{tabular}
\end{minipage}
\end{table*}

\section{Summary}

\begin{enumerate}
      \item High-resolution, deep, near-infrared ($JHK_S$) images of the region towards IRAS~06345-3023 reveals a small group of stars compatible with the presence of a small aggregate of young stars. 
      \item The stars are located in a region exhibiting nebular emission and are embedded in a dense molecular cloud core detected through CO line emission. 
      \item The nebular light is rich in morphological structures, including arcs in the vicinity of embedded stars, wisps, and bright rims of a butterfly-shaped dark cloud. In particular, the circular arc may represent a circumbinary ring.
      \item At about 1,5 kpc (heliocentric distance) in the outer Galaxy, and at a galactocentric distance of 10~kpc, the relatively large value of its galactic latitude places this group of young stars at 450~pc below the Galactic plane, at the edge of the molecular disc. Thus, active star formation is taking place at vertical distances larger than those typical of the (thin) disc. This may raise questions on the origin of the gas forming this cloud and on the sharpness of the stellar and of the gaseous Galactic thin disc.
      \item The group of young stars is likely to be composed of low-mass stars, mostly Class~I and flat spectrum young stellar objects.
      \item Inspection of WISE images and catalogues confirms the YSO nature of the sources and the absence of additional young stars associated to this star formation site.

   \end{enumerate}

\section*{Acknowledgments}

This research was based on observations collected at the ESO 8.2-m VLT-UT1 Antu telescope (program 68.C-0214A). JY acknowledges support from FCT (SFRH/BSAB/1423/2014).
This research made use of the NASA/ IPAC Infrared Science Archive, which is operated by the Jet Propulsion Laboratory, California Institute of Technology, under contract with the National Aeronautics and Space Administration.
This research also made use of the SIMBAD database, operated at CDS, Strasbourg, France, as well as SAOImage DS9, developed by the Smithsonian Astrophysical Observatory.
This publication makes use of data products from the Wide-field Infrared Survey Explorer, which is a joint project of the University of California, Los Angeles, and the Jet Propulsion Laboratory/California Institute of Technology, funded by the National Aeronautics and Space Administration.

\label{lastpage}

\end{document}